\documentclass{article}
\begin{document}

\title{The motion equation for massive fermions and bosons without Higgs}
\author{Gunn Quznetsov \\
%EndAName
\bigskip 
454016, Chelyabinsk-16, yD.BET. $\Phi $ N 949892, Russia\\
email: lak@cgu.chel.su, gunn@mail.ru, quznets@yahoo.com}
\date{}
\maketitle

\begin{abstract}
The motion equation with nonzero mass, invariant for weak isospin
transformation, can be obtained from the Dirac equation by the adding of the
Clifford pentad fifth element. The motion equation of the SU(2) Yang-Mills
field components is similar to the Klein-Gordon equation with the nonzero
mass.
\end{abstract}

PACS: 13.10, 12.10, 12.15, 14.70.Fm, 14.80.B, 14.80.C \bigskip

\section{Introduction}

I consider the motion equations for fermions with nonzero masses without Higgs 
with all five elements of Clifford's pentad. And i reform the motion equation 
for components W-field as equation, similar to the nonzero Klein-Gordon equation.

\bigskip 
Notations:

\begin{eqnarray*}
&&1_2\stackrel{Def}{=}\left[ 
\begin{array}{cc}
1 & 0 \\ 
0 & 1
\end{array}
\right] \mbox{, }1_4\stackrel{Def}{=}\left[ 
\begin{array}{cc}
1_2 & 0_2 \\ 
0_2 & 1_2
\end{array}
\right] \mbox{, } \\
&&0_2\stackrel{Def}{=}\left[ 
\begin{array}{cc}
0 & 0 \\ 
0 & 0
\end{array}
\right] \mbox{, }0_4\stackrel{Def}{=}\left[ 
\begin{array}{cc}
0_2 & 0_2 \\ 
0_2 & 0_2
\end{array}
\right]
\end{eqnarray*}

\[
\gamma ^{\left[ 5\right] }\stackrel{Def}{=}\left[ 
\begin{array}{cc}
1_2 & 0_2 \\ 
0_2 & -1_2
\end{array}
\right] \mbox{, }\beta ^{\left[ 0\right] }\stackrel{Def}{=}-1_4\mbox{, }1_8%
\stackrel{Def}{=}\left[ 
\begin{array}{cc}
1_4 & 0_4 \\ 
0_4 & 1_4
\end{array}
\right] \mbox{,} 
\]

the Pauli matrices:

\[
\sigma _1=\left[ 
\begin{array}{cc}
0 & 1 \\ 
1 & 0
\end{array}
\right] \mbox{, }\sigma _2=\left[ 
\begin{array}{cc}
0 & -\mathrm{i} \\ 
\mathrm{i} & 0
\end{array}
\right] \mbox{, }\sigma _3=\left[ 
\begin{array}{cc}
1 & 0 \\ 
0 & -1
\end{array}
\right] \mbox{;} 
\]

if $\vartheta $ is the $4\times 4$ matrix then

\[
\underline{\vartheta }\stackrel{Def}{=}\left[ 
\begin{array}{cc}
\vartheta & 0_4 \\ 
0_4 & \vartheta
\end{array}
\right] \mbox{.} 
\]

A set $\widetilde{C}$ of complex $n\times n$ matrices is denoted as {\it %
Clifford's set} \cite{Md} if the following conditions are fulfilled:

if $\alpha _k\in \widetilde{C}$ and $\alpha _r\in \widetilde{C}$ then $%
\alpha _k\alpha _r+\alpha _r\alpha _k=2\delta _{k,r}$;

if $\alpha _k\alpha _r+\alpha _r\alpha _k=2\delta _{k,r}$ for all elements $%
\alpha _r$ of $\widetilde{C}$ then $\alpha _k\in \widetilde{C}$.

If $n=4$ then Clifford's set either contains $3$ matrices ({\it %
Clifford's triplet}) or contains $5$ matrices ({\it Clifford's pentad}).

For example: {\it Clifford pentad} $\beta $:

\begin{equation}
\beta ^{\left[ 1\right] }\stackrel{Def}{=}\left[ 
\begin{array}{cc}
\sigma _1 & 0_2 \\ 
0_2 & -\sigma _1
\end{array}
\right] \mbox{, }\beta ^{\left[ 2\right] }\stackrel{Def}{=}\left[ 
\begin{array}{cc}
\sigma _2 & 0_2 \\ 
0_2 & -\sigma _2
\end{array}
\right] \mbox{, }\beta ^{\left[ 3\right] }\stackrel{Def}{=}\left[ 
\begin{array}{cc}
\sigma _3 & 0_2 \\ 
0_2 & -\sigma _3
\end{array}
\right] \mbox{,}  \label{lghr}
\end{equation}

\begin{equation}
\gamma ^{\left[ 0\right] }\stackrel{Def}{=}\left[ 
\begin{array}{cc}
0_2 & 1_2 \\ 
1_2 & 0_2
\end{array}
\right] \mbox{, }  \label{lghr1}
\end{equation}

\begin{equation}
\beta ^{\left[ 4\right] }\stackrel{Def}{=}\mathrm{i}\left[ 
\begin{array}{cc}
0_2 & 1_2 \\ 
-1_2 & 0_2
\end{array}
\right] .  \label{lghr2}
\end{equation}

\section{Fermions}

Let the motion equation (Dirac equation) for the free ferm{i}on be in the
following form:

\begin{equation}
\left( \beta ^{\left[ 0\right] }\mathrm{i}\partial _0+\beta ^{\left[
1\right] }\mathrm{i}\partial _1+\beta ^{\left[ 2\right] }\mathrm{i}\partial
_2+\beta ^{\left[ 3\right] }\mathrm{i}\partial _3+\gamma ^{\left[ 0\right]
}m\right) \psi =0\mbox{.}  \label{e}
\end{equation}

The motion equation for two independent fermions with masses $m_1$ and $m_2$
(a two-fermion state) is of the following:

\[
\left( \underline{\beta ^{\left[ 0\right] }}\mathrm{i}\partial _0+\underline{%
\beta ^{\left[ 1\right] }}\mathrm{i}\partial _1+\underline{\beta ^{\left[
2\right] }}\mathrm{i}\partial _2+\underline{\beta ^{\left[ 3\right] }}%
\mathrm{i}\partial _3+\underline{\gamma ^{\left[ 0\right] }}m_{1,2}\right)
\Psi =0 
\]

with

\[
m_{1,2}\stackrel{Def}{=}\left[ 
\begin{array}{cc}
m_1\cdot 1_4 & 0_4 \\ 
0_4 & m_2\cdot 1_4
\end{array}
\right] \mbox{.} 
\]

This equation is not invariant for the SU(2) isospin transformation

\[
U\stackrel{Def}{=}\left[ 
\begin{array}{cccc}
\left( a+\mathrm{i}b\right) 1_2 & 0_2 & \left( c+\mathrm{i}q\right) 1_2 & 0_2
\\ 
0_2 & 1_2 & 0_2 & 0_2 \\ 
\left( -c+\mathrm{i}q\right) 1_2 & 0_2 & \left( a-\mathrm{i}b\right) 1_2 & 
0_2 \\ 
0_2 & 0_2 & 0_2 & 1_2
\end{array}
\right] 
\]

with

\[
a^2+b^2+c^2+q^2=1 
\]

because it holds four, only, elements of the Clifford pentad.

If $\partial _\mu U=U\partial _\mu $ then the equation with all five elements

\[
\left( \underline{\beta ^{\left[ 0\right] }}\mathrm{i}\partial _0+\underline{%
\beta ^{\left[ 1\right] }}\mathrm{i}\partial _1+\underline{\beta ^{\left[
2\right] }}\mathrm{i}\partial _2+\underline{\beta ^{\left[ 3\right] }}%
\mathrm{i}\partial _3+\underline{\gamma ^{\left[ 0\right] }}m_5+\underline{%
\beta ^{\left[ 4\right] }}m_4\right) \Psi =0 
\]

is invariant for the following transformation:

\begin{eqnarray*}
\Psi &\rightarrow &\Psi ^{\prime }=U\Psi \mbox{;} \\
m_5 &\rightarrow &m_5^{\prime }=am_5\left( \ell _{\circ }+\ell _{\circ
}\right) -\sqrt{1-a^2}m_4\left( \ell _{\circ }-\ell _{\circ }\right) \mbox{;}
\\
m_4 &\rightarrow &m_4^{\prime }=am_4\left( \ell _{\circ }+\ell _{\circ
}\right) +\sqrt{1-a^2}m_5\left( \ell _{\circ }-\ell _{\circ }\right)
\end{eqnarray*}

with

\[
\ell _{\circ }\stackrel{Def}{=}\frac 1{2\sqrt{\left( 1-a^2\right) }}\left[ 
\begin{array}{cc}
\left( b+\sqrt{\left( 1-a^2\right) }\right) 1_4 & \left( q-\mathrm{i}%
c\right) 1_4 \\ 
\left( q+\mathrm{i}c\right) 1_4 & \left( \sqrt{\left( 1-a^2\right) }%
-b\right) 1_4
\end{array}
\right] \mbox{,} 
\]

\[
\ell _{*}\stackrel{Def}{=}\frac 1{2\sqrt{\left( 1-a^2\right) }}\left[ 
\begin{array}{cc}
\left( \sqrt{\left( 1-a^2\right) }-b\right) 1_4 & \left( -q+\mathrm{i}%
c\right) 1_4 \\ 
\left( -q-\mathrm{i}c\right) 1_4 & \left( b+\sqrt{\left( 1-a^2\right) }%
\right) 1_4
\end{array}
\right] \mbox{.}
\]

In this transformation the value of $m_5^2+m_4^2$ is invariant.

Hence we have got the two-fermion motion equation, invariant for global weak
isospin transformation, with nonzero nonHiggs mass.

If $g$ is a real number, $W_\mu ^{0,}$, $W_\mu ^{1,}$,$W_\mu ^{2,}$ are
real scalar functions on the 3+1 space-time and

\[
W_\mu \stackrel{Def}{=} \left[ 
\begin{array}{cccc}
W_\mu ^{0,}1_2 & 0_2 & \left( W_\mu ^{1,}-\mathrm{i}W_\mu ^{2,}\right) 1_2 & 
0_2 \\ 
0_2 & 1_2 & 0_2 & 0_2 \\ 
\left( W_\mu ^{1,}+\mathrm{i}W_\mu ^{2,}\right) 1_2 & 0_2 & -W_\mu ^{0,}1_2
& 0_2 \\ 
0_2 & 0_2 & 0_2 & 1_2
\end{array}
\right] 
\]

then the equation

\[
\left( \sum_{\mu =0}^3\underline{\beta ^{\left[ \mu \right] }}\mathrm{i}%
\left( \partial _\mu -\mathrm{i}\frac 12gW_\mu \right) +\underline{\gamma
^{\left[ 0\right] }}m_5+\underline{\beta ^{\left[ 4\right] }}m_4\right) \Psi
=0 
\]

is invariant for the local weak isospin transformation:

\begin{eqnarray*}
\Psi &\rightarrow &\Psi ^{\prime }=U\Psi \mbox{;} \\
m_5 &\rightarrow &m_5^{\prime }=am_5\left( \ell _{\circ }+\ell _{*}\right) -%
\sqrt{1-a^2}m_4\left( \ell _{\circ }-\ell _{*}\right) \mbox{;} \\
m_4 &\rightarrow &m_4^{\prime }=am_4\left( \ell _{\circ }+\ell _{*}\right) +%
\sqrt{1-a^2}m_5\left( \ell _{\circ }-\ell _{*}\right) \mbox{,} \\
W_\mu &\rightarrow &W_\mu ^{\prime }=UW_\mu U^{\dagger }-\frac{2\mathrm{i}}{%
g}\left( \partial _\mu U\right) U^{\dagger }\mbox{.}
\end{eqnarray*}

\section{Bosons}

The motion equation of the Yang-Mills SU(2) field in the space without
matter (for instance \cite{Sd} or \cite{Rd} ) has got the following form:

\[
\partial ^\nu \mathbf{W}_{\mu \nu }=-g\mathbf{W}^\nu \times \mathbf{W}_{\mu
\nu } 
\]

with:

\[
\mathbf{W}_{\mu \nu }=\partial _\mu \mathbf{W}_\nu -\partial _\nu \mathbf{W}%
_\mu +g\mathbf{W}_\mu \times \mathbf{W}_\nu 
\]

and

\[
\mathbf{W}_\mu =\left[ 
\begin{array}{c}
W_\mu ^{0,} \\ 
W_\mu ^{1,} \\ 
W_\mu ^{2,}
\end{array}
\right] \mbox{.} 
\]

Hence the motion equation for $W_\mu ^{0,}$ is the following:

\begin{equation}
\begin{array}{c}
\partial ^\nu \partial _\nu W_\mu ^{0,}=g^2\left( W^{2,\nu }W_\nu
^{2,}+W^{1,\nu }W_\nu ^{1,}\right) W_\mu ^{0,}- \\ 
-g^2\left( W^{1,\nu }W_\mu ^{1,}+W^{2,\nu }W_\mu ^{2,}\right) W_\nu ^{0,}+
\\ 
+g\partial ^\nu \left( W_\mu ^{1,}W_\nu ^{2,}-W_\mu ^{2,}W_\nu ^{1,}\right) +
\\ 
+g\left( W^{1,\nu }\partial _\mu W_\nu ^{2,}-W^{1,\nu }\partial _\nu W_\mu
^{2,}-W^{2,\nu }\partial _\mu W_\nu ^{1,}+W^{2,\nu }\partial _\nu W_\mu
^{1,}\right) + \\ 
+\partial ^\nu \partial _\mu W_\nu ^{0,}\mbox{.}
\end{array}
\label{b}
\end{equation}

$W_\mu ^{1,}$ and $W_\mu ^{2,}$ satisfy to similar equations.

This equation can be reformed as the following:

\[
\begin{array}{c}
\partial ^\nu \partial _\nu W_\mu ^{0,}=\left[ g^2\left( W^{2,\nu }W_\nu
^{2,}+W^{1,\nu }W_\nu ^{1,}+W^{0,\nu }W_\nu ^{0,}\right) \right] \cdot W_\mu
^{0,}- \\ 
-g^2\left( W^{1,\nu }W_\mu ^{1,}+W^{2,\nu }W_\mu ^{2,}+W^{0,\nu }W_\mu
^{0,}\right) W_\nu ^{0,}+ \\ 
+g\partial ^\nu \left( W_\mu ^{1,}W_\nu ^{2,}-W_\mu ^{2,}W_\nu ^{1,}\right) +
\\ 
+g\left( W^{1,\nu }\partial _\mu W_\nu ^{2,}-W^{1,\nu }\partial _\nu W_\mu
^{2,}-W^{2,\nu }\partial _\mu W_\nu ^{1,}+W^{2,\nu }\partial _\nu W_\mu
^{1,}\right) + \\ 
+\partial ^\nu \partial _\mu W_\nu ^{0,}\mbox{.}
\end{array}
\]

This equation looks like to the Klein-Gordon equation of field $W_\mu ^{0,}$
with mass

\begin{equation}
g\left[ -\left( W^{2,\nu }W_\nu ^{2,}+W^{1,\nu }W_\nu ^{1,}+W^{0,\nu }W_\nu
^{0,}\right) \right] ^{\frac 12}.  \label{z10}
\end{equation}

and with the additional terms of the $W_\mu ^{0,}$ interactions with others
components of $\mathbf{W}$.

''Mass'' (\ref{z10}) is invariant for the following transformations:

\[
\left\{ 
\begin{array}{c}
W_r^{k,\prime }=W_r^{k,}\cos \lambda -W_s^{k,}\sin \lambda \mbox{.} \\ 
W_s^{k,\prime }=W_r^{k,}\sin \lambda +W_s^{k,}\cos \lambda \mbox{;}
\end{array}
\right| 
\]

\[
\left\{ 
\begin{array}{c}
W_0^{k,\prime }=W_0^{k,}\cosh \lambda -W_s^{k,}\sinh \lambda \mbox{,} \\ 
W_s^{k,\prime }=W_s^{k,}\cosh \lambda -W_0^{k,}\sinh \lambda
\end{array}
\right| 
\]

with a real number $\lambda $, and $r\in \left\{ 1,2,3\right\} $, and $s\in
\left\{ 1,2,3\right\} $,

and (\ref{z10}) is invariant for a global weak isospin transformation $U$:

\[
W_\nu \rightarrow W_\nu ^{\prime }=UW_\nu U^{\dagger } 
\]

but is not invariant for a local transformation $U$:

\[
W_\nu \rightarrow W_\nu ^{\prime }=UW_\nu U^{\dagger }-\frac{2\mathrm{i}}%
g\left( \partial _\nu U\right) U^{\dagger }\mbox{.} 
\]

Equation (\ref{b}) can be simplified as follows:

\[
\begin{array}{c}
\sum_\nu g_{\nu ,\nu }\partial ^\nu \partial _\nu W_\mu ^{0,}=\left[
g^2\sum_{\nu \neq \mu }g_{\nu ,\nu }\left( \left( W_\nu ^{2,}\right)
^2+\left( W_\nu ^{1,}\right) ^2\right) \right] \cdot W_\mu ^{0,}- \\ 
-g^2\sum_{\nu \neq \mu }g_{\nu ,\nu }\left( W^{1,\nu }W_\mu ^{1,}+W^{2,\nu
}W_\mu ^{2,}\right) W_\nu ^{0,}- \\ 
+g\sum_\nu g_{\nu ,\nu }\partial ^\nu \left( W_\mu ^{1,}W_\nu ^{2,}-W_\mu
^{2,}W_\nu ^{1,}\right) + \\ 
+g\sum_\nu g_{\nu ,\nu }\left( W^{1,\nu }\partial _\mu W_\nu ^{2,}-W^{1,\nu
}\partial _\nu W_\mu ^{2,}-W^{2,\nu }\partial _\mu W_\nu ^{1,}+W^{2,\nu
}\partial _\nu W_\mu ^{1,}\right) + \\ 
+\partial _\mu \sum_\nu g_{\nu ,\nu }\partial ^\nu W_\nu ^{0,}\mbox{.}
\end{array}
\]

(here no of summation over indexes $_\nu ^\nu $; the summation is expressed
by $\sum $ ) with $g_{0,0}=1$, $g_{1,1}=g_{2,2}=g_{3,3}=-1$.

In this equation the form

\[
g\left[ -\sum_{\nu \neq \mu }g_{\nu ,\nu }\left( \left( W_\nu ^{2,}\right)
^2+\left( W_\nu ^{1,}\right) ^2\right) \right] ^{\frac 12} 
\]

varies in space, but it does not contain $W_\mu ^{0,}$ and locally acts as
mass - i.e. it does not allow to particles of this field to behave as a
massless ones.

Hence if Higgs exist then mass, calculated with Higgs, should be corrected
with this ''mass''. If Higgs do not exist then this ''mass'' turns out to be
the only reason, giving a $W$-boson nonzero mass in measurements \cite{Kla}.

Therefore the $W$ fields should behave as the massive fields with Higgs or
without Higgs.

\section{Conclusion}

Therefore Higgs are not necessary for the nonzero leptons' and bosons' masses.

\end{document}